%% file: manuscript.tex
\documentclass[%
superscriptaddress,
showpacs,preprintnumbers,
nofootinbib,
nobibnotes,
bibnotes,
 amsmath,amssymb,
 aps,
 prstab, twocolumn,
]{revtex4-1}
\usepackage{tabularx}
\usepackage{textcomp}
\usepackage{graphicx}%
\usepackage{framed}
\usepackage[dvipsnames]{xcolor}
\usepackage{appendix}
\usepackage{amsmath}
\usepackage{amssymb}
\usepackage{dcolumn}
\usepackage{graphicx}
\usepackage{longtable}
\usepackage{bm}
\usepackage{color}
\usepackage{epsfig}
\usepackage[colorlinks, citecolor=blue, urlcolor=black, bookmarks=false, linkcolor=blue, hypertexnames=true]{hyperref}

\usepackage{hyperref}
\hypersetup{
    colorlinks=true,
    linkcolor=blue,
    filecolor=magenta,      
    urlcolor=blue,
    pdftitle={Overleaf Example},
    pdfpagemode=FullScreen,
    }

\begin{document}

\title{Defect-induced localization of information in 1D Kitaev model}
\affiliation{Department of Physics, Purdue University, West Lafayette, Indiana 47907, USA}
\affiliation{Department of Chemistry, Purdue University, West Lafayette, Indiana 47907, USA}
\affiliation{Purdue Quantum Science and Engineering Institute, Purdue University, West Lafayette, Indiana 47907, USA}
\affiliation{Elmore Family School of Electrical \& Computer Engineering, Purdue University, West Lafayette, Indiana 47907, USA}

\author{Varadharajan Muruganandam}
\affiliation{Department of Physics, Purdue University, West Lafayette, Indiana 47907, USA}
\affiliation{Purdue Quantum Science and Engineering Institute, Purdue University, West Lafayette, Indiana 47907, USA}

\author{Manas Sajjan}
 \affiliation{Department of Chemistry, Purdue University, West Lafayette, Indiana 47907, USA}
 \affiliation{Purdue Quantum Science and Engineering Institute, Purdue University, West Lafayette, Indiana 47907, USA}

\author{Sabre Kais}
\email{kais@purdue.edu}
\affiliation{Department of Physics, Purdue University, West Lafayette, Indiana 47907, USA}
\affiliation{Department of Chemistry, Purdue University, West Lafayette, Indiana 47907, USA}
\affiliation{Purdue Quantum Science and Engineering Institute, Purdue University, West Lafayette, Indiana 47907, USA}
\affiliation{Elmore Family School of Electrical \& Computer Engineering, Purdue University, West Lafayette, Indiana 47907, USA}

\begin{abstract}
We discuss one-dimensional(1D) spin compass model or 1D Kitaev model in the presence of local bond defects. Three types of local disorders concerning both bond-nature and bond-strength that occur on kitaev materials have been investigated. Using exact diagonalization, two-point spin-spin structural correlations and four-point Out-of-Time-Order Correlators(OTOC) have been computed for the defective spin chains. The proposed quantities give signatures of these defects in terms of their responses to location and strength of defects. A key observation is that the information in the OTOC space gets trapped at the defect site giving rise to the phenomena of localization of information thus making these correlators a suitable diagnostic tool to detect and  characterize these defects.
\end{abstract}
\maketitle
\section{Introduction}

Spin Compass Models(SCM)\cite{RevModPhys.87.1} are spin models with Ising-type interactions along directions that are dependent on bond directions. A well-known SCM is the Kitaev's Honeycomb spin model which exhibits a quantum spin liquid (QSL)phase\cite{Kitaev2006anyons} supporting abelian and non-abelian anyonic excitations. This model has an exotic phase diagram with rich topological properties that offer the promise of fault-tolerant quantum computation. Along the materials side, with the recent blow-up of both theoretical and experimental studies of the iridium-oxide materials \cite{Kitaev1}, the $\alpha-RuCl_3$\cite{koitzsch2016j} has garnered enormous attention. Particularly, neutron scattering \cite{banerjee2016proximate,banerjee2017neutron} and thermal conductivity\cite{cond1,cond2} experiments have provided evidence that the Kitaev-type interactions dominate the physics of $\alpha-RuCl_3$ thus making them suitable candidate materials for realizing the Kitaev model. One of the main barriers in filling the gap between the theoretical predictions and real materials is the presence of defects and disorders. \newline
\par Defects in real materials change the physical properties of the system that usually do not have a counterpart in their clean limit. Particularly, disorders like vacancies, impurities and lattice distortions that are inevitable in these materials contribute to instabilities\cite{andrade2020susceptibility}, divergences in their density of states\cite{PhysRevLett.122.047202}  and localization effects\cite{willans2010disorder,kao2021disorder}. On the other hand, such defects can also open up a plethora of new phases with unpaired majorana modes\cite{unpaired1,unpaired2} that arise as twist defects as proposed by Bombin\cite{bombin}. These defects being the epicenter of these modes show braiding statistics that are tolerant to local perturbations. Recently, this phenomena has been generalized to arbitrary tri-valent planar lattices with Kitaev-type interactions\cite{yan2023generalized}. Pertaining to these reasons, the study of defects on pristine models becomes an essential venture as a part of theoretical analysis of the aforementioned materials. Towards this direction, as a first step, we study in this article various kinds of defects on the one-dimensional(1D) analog of 2D Kitaev model i.e. 1D compass model and uncover characteristics of the system using structural and dynamical quantities. The signatures observed in these quantities, as we shall show in the following sections serve as  diagnostic tools to detect, observe and characteize the considered defects in real systems. \newline
\par
The paper is organized as follows: In Sec. \ref{sec1}, we introduce the model, type of disorders and describe the involved metrics and numerical methods. In Sec. \ref{sec2}, we present the results of disorder effects on both structural and dynamical properties of the ground state by computing different correlation measures that are introduced in the previous section. Section \ref{sec3} is the conclusion.

\section{Models}\label{sec1}
\subsection*{Kitaev Models}
\begin{figure*}[!thb]
\begin{framed}
\centerline{
\includegraphics[width = 6 in]{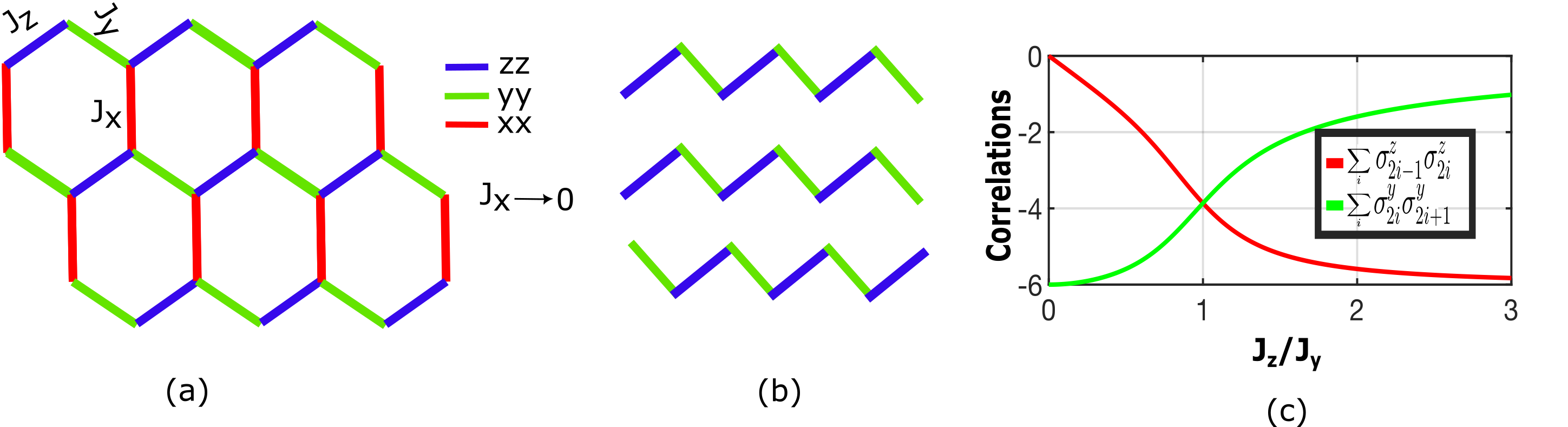}}
\caption{
(a) 2D Kitaev model on a honeycomb model
(b) decoupled 1D Kitaev chains 
(c) correlations for the clean limit of 1D Kitaev model for L=12 spins under periodic boundary conditions(PBC)
}\label{fig:Figure_1}
\end{framed}
\end{figure*}

Kitaev model in 2D is a bond-dependent interacting spin graph as shown in Fig. \ref{fig:Figure_1}(a) given by the Hamiltonian,
\begin{align}
    H_{\text{2D}}=& J_x\sum_{x-bonds}\sigma_{i}^x\sigma_{j}^x+J_y\sum_{y-bonds}\nonumber\\&\sigma_{i}^y\sigma_{j}^y+J_z\sum_{z-bonds}\sigma_{i}^z\sigma_{j}^z
\end{align}
The above system belongs to the larger umbrella of SCMs. Such systems can also be realized on arbitrary trivalent graphs like square-octagon lattice\cite{sq_oct1,sq_oct3,yamada1,yamada2} within cyclooctatraene based polymeric platforms\cite{varadharajan2023}. Recent studies have shown that the 2D Kitaev lattice can be approximated by coupled 1D SCM chains\cite{Kiteav1D_1,Kiteav1D_2,feng2023dimensional} as shown in Fig. \ref{fig:Figure_1} and show interesting similarities\cite{adhip2023} in terms of its phase diagram and many other physical properties. This calls us to give extra attention to the one-dimensional(1D) ZY SCM\cite{1dcompass,1dcompass2} based on our convention. The 1D model is a bond-alternating spin-1/2 chain with bond-dependent ZZ and YY interactions as shown in \ref{fig:Figure_1}(b). The Hamiltonian is given by, 

\begin{align}
H_{\text{1D}}=&J_z\sum_{i=1}^{N/2}\sigma_{2i-1}^z \sigma_{2i}^z+J_y\sum_{i=1}^{N/2}\sigma_{2i}^y \sigma_{2i+1}^y\label{1D}
\end{align}
where $J_y,\,J_z$ are the alternating bond strengths of y-bonds and z-bonds respectively. $N$ typically denotes the number of unit cells. 
\subsection*{1. Clean limit}
For the clean limit, the Kiteav model in 1D undergoes a continuous quantum phase transition(QPT) from a phase with dominating  zz correlations on odd bonds for $Jz/Jy<1$ to a phase with dominating yy correlations on even bonds for $Jz/Jy>1$ as shown in \ref{fig:Figure_1}(c) with transition point at $J_z=J_y$. We consider the following defects inspired by twist and on-site disorders that occur in 2D QSL model and show that in its 1D limit, the structural correlations and dynamic OTOCs  can give signatures of these defects. For the sake of convenience, we have considered the ZY model and the results obtained in this article are general and remain same for XY and XZ models as well. 
\subsection*{2. Defects}\label{sec2}
\begin{figure}[h]
\centerline{
\includegraphics[width=3.3 in]{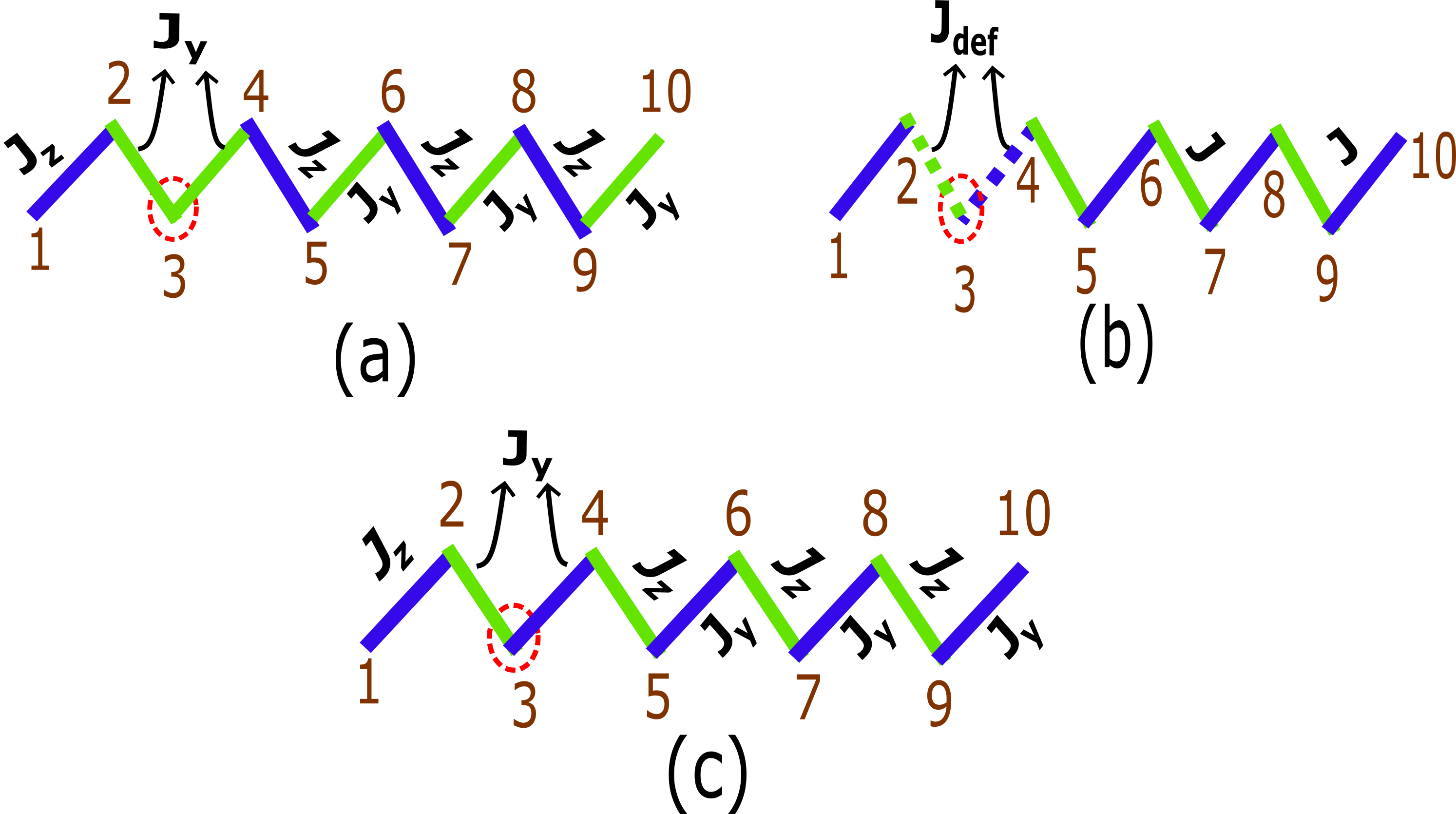}}
\caption{
(a) Type 1: Bond nature-flip defect at an odd defect site 3 with yy coupling
(b) Type 2: Bond-strength defect at defect site 3 with $J_{\text{def}}$ being the bond strength of disordered bonds and $J$ the bond strength of other bonds.
(c) Type 3: Bond strength-flip defect at defect site 3 with $J_y$ and $J_z$ being the bond strengths as shown. 
}\label{fig:Figure_2}
\end{figure}

The kind of defects that we examine in this paper are local defects that occur on a particular site concerning its local bonds. These defects are different from the usual disorders that are either taken to occur at every site or at every nearest neighbour interaction (i.e. bond) that are commonly studied in spin-chains.
Bond flip and bond strength defects appear both in 1D and 2D Kitaev Hamiltonians ubiquitously where the alternating structure is broken at the defect site. Firstly, we consider defect of type 1 wherein the bond nature at the vicinity of the defect site is flipped and repeated say for instance, the repeating yy bond at defect site 3 in Fig: \ref{fig:Figure_2}(a) and the alternating nature is preserved before and after the defect. The second type concerns the bond strengths wherein the bonds at the defect have a weaker bond strength compared to the bonds elsewhere as in Fig: \ref{fig:Figure_2}(b).The third type concerns the bond strength at the vicinity of the defect site being flipped and repeated say for instance, the repeating Jy-Jy bond strengths at defect site 3 in Fig: \ref{fig:Figure_2}(c). An additional motivation of studying these defects on a 1D model is the appearance of these defects as effective 1D line defects\cite{linedefect} on 2D QSL. 

\section{Metrics and methods}\label{sec2}
\subsection*{Spin-Spin correlations}  
We compute $\sigma_i^z\sigma_j^z$ and $\sigma_i^y\sigma_j^y$ defined as in Eq. \ref{eq_corr}, correlations for spin chain  upto L=12 spins based on Exact diagonalization by employing periodic boundary condtions(PBC). Further, we compare these correlation plots with the clean limit (Fig.\ref{fig:Figure_1}(c)) and look for signatures for these defects in terms of their structural correlation measures. 
\begin{align}
&\langle\sigma_{2i-1}^z\sigma_{2i}^z\rangle=\sum_i\langle\psi_{\text{g}}|\sigma_{2i-1}^z\sigma_{2i}^z|\psi_{\text{g}}\rangle\nonumber\\
&\langle\sigma_{2i}^y\sigma_{2i+1}^y\rangle=\sum_i\langle\psi_{\text{g}}|\sigma_{2i}^y\sigma_{2i+1}^y|\psi_{\text{g}}\rangle\ \label{eq_corr}  
\end{align}
where $\psi_g$ denotes the ground state of the considered defective spin chain types. 
\subsection*{Out-of-Time-Order Correlator(OTOC)}  
\begin{figure}[!thb]
\centerline{
\includegraphics[width = 2 in]{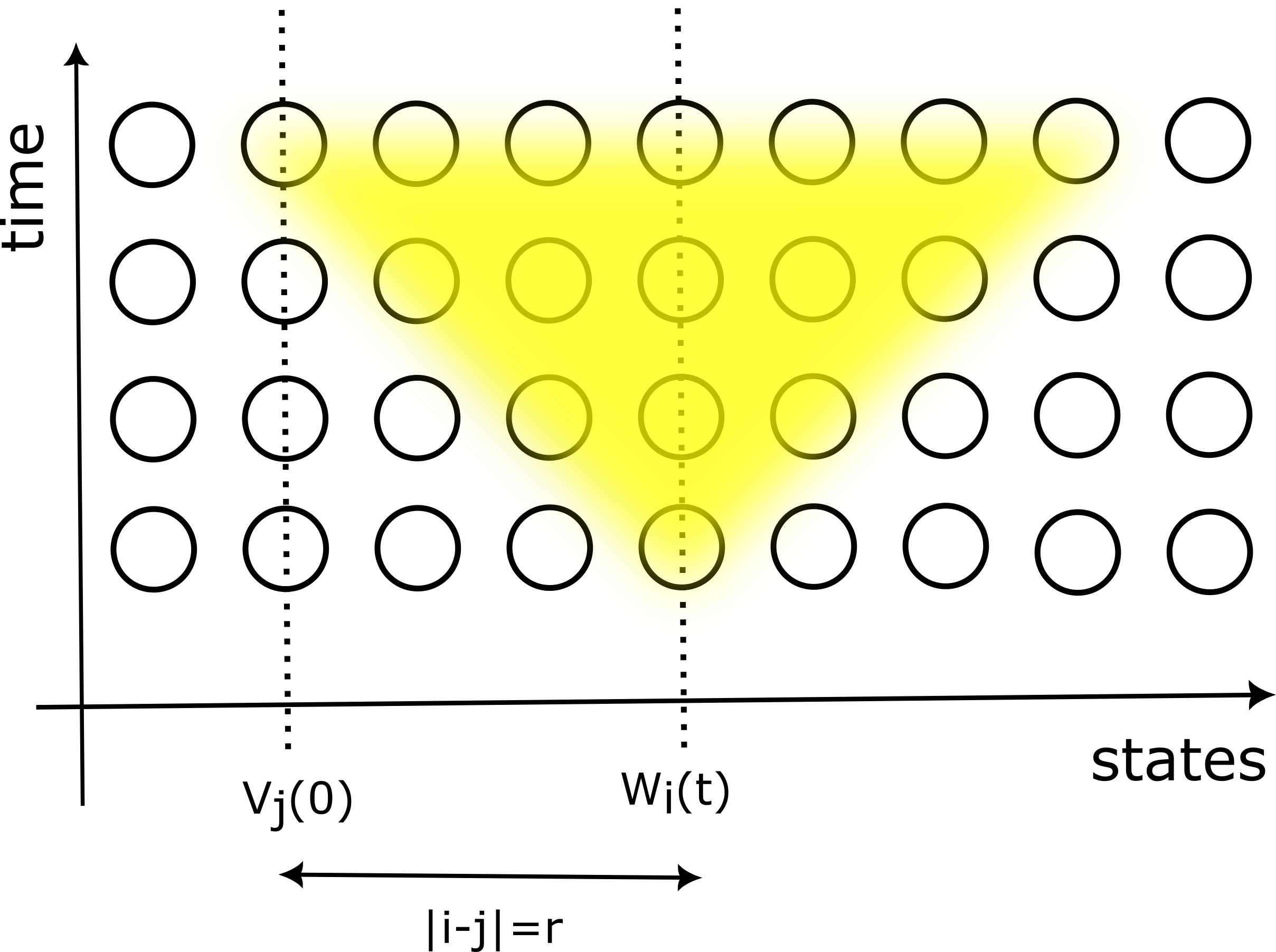}}
\caption{
space-time propagation of a general OTOC $F(r,t)$ signal
}\label{fig:Figure_3}
\end{figure}

Out-of-Time-Order Correlator(OTOC) first introduced by Larkin and Ovchinnikov in the context of superconductivity\cite{larkin} has been exploited as a tool to provide interesting insights into physical systems. Most considerably, OTOC being a dynamic quantity quantifies how local information belonging to local degrees of freedom and operators spreads across global degrees of freedom of a quantum many-body system which is typically inaccessible to local probes. This classifies OTOC as an information measure that gives information about the scrambling dynamics of the considered physical model. Further, OTOC has found applications in the field of quantum chaos ranging from condensed matter\cite{cond1,otoc_luttinger,ising_otoc} to high-energy physics. The key idea is the connection between the growth exponent of OTOC called the butterfly velocity of information and Lyapunov exponent indicating the onset of chaos.\cite{Maldacena,shenker,gu2017OTOC}. Moreover, Recent proposals have shown that OTOC serves as a useful quantity to detect phase transitions such as  Many-Body Localization(MBL) \cite{riddell_mbl,mbl_otoc}and dynamical phase transitions\cite{ising_otoc,exp_dqpt,floquet_dpt}.
The OTOC is a 4-point correlation measure defined as $F_{WV}^{i,j}(t) = \langle W_i(t)V_j(0)W_i(t)V_j(0)\rangle$ where $(W,V)$ are local operators for sites $(i,j)$ computed at time $(t,0)$ respectively. Further, The OTOC signal with respect to this 4-point correlation function can be defined as $F(r=|i-j|,t) = \langle W_i(t)V_j(0)W_i(t)V_j(0)\rangle$ whose space-time propagation as shown in Fig. \ref{fig:Figure_3} can give insight into information scarmbling dynamics of the system thus making it a suitable probe for information propogation. OTOCs by itself is a complex quantity with the real part related to the squared commutator $C_{i,j}(t) = \langle[W_{i}(t),V_{j}(0)]^+[W_{i}(t),V_{j}(0)]\rangle=\langle|[W_{i}(t),V_{j}(0)]|^2\rangle=2(1- \text{Re}[F_{i,j}(t)])$. Recently, It has been shown that the imaginary part of OTOC also possesses interesting properties of the physical system \cite{manas_otoc} that in turn is given by both commutators as well as anti-commutator of the local operators $W\,\, \text{\&}\,\,V$ respectively. The real and imaginary of the OTOC is associated to the commutator and anti-commutator as follows,
\begin{align}
\langle|[W_{i}(t),V_{j}(0)]|^2\rangle=2(1- \text{Re}[F_{WV}(r,t)]) \label{eq_real_otoc}
\end{align}
\begin{align}
\langle[W_{i}(t),V_{j}(0)]^+\{W_{i}(t),V_{j}(0)\}\rangle=2(1- \text{Im}[F_{WV}(r,t)])  \label{eq_imag_otoc}
\end{align}
Since both the real and imaginary parts of OTOC in Eqs. \ref{eq_real_otoc}-\ref{eq_imag_otoc} contain the common commutator, evaluation of the commutator becomes necessary. The dynamics of OTOC is controlled by the Heisenberg evolutions of $W_i(t)$ and analytical expressions for the same could be in principle derived using Baker-Campbell-Haunsdroff(BCH) expansion\cite{hall2013lie}. For our model, we choose the operator XX($\sigma_x^i\sigma_x^j$) as the OTOC operator based on our observations that other YY and ZZ OTOCs do not capture the information about the defects. Further, to see how information spreads from one end of the chain to its other end in the presence of aforementioned defects, we fix $i=1$ and vary $j=1, 2, 3......L$ thus making our OTOC probe take the form, $\langle \sigma_1^x(t)\sigma_j^x(0)\sigma_1^x(t)\sigma_j^x(0)\rangle$ defined as in Eq. \ref{eq_otoc}. We compute the OTOC signal by means of Exact Diagonalization(ED) for spin chains upto L=12 spins under open boundary conditions (OBC). 
\begin{align}
\langle \sigma_1^x(t)\sigma_j^x(0)\sigma_1^x(t)\sigma_j^x(0)\rangle=\langle\psi_g| \sigma_1^x(t)\sigma_j^x(0)\sigma_1^x(t)\sigma_j^x(0)|\psi_g\rangle \label{eq_otoc}  
\end{align} 
where $\psi_g$ denotes the ground state of the considered defective spin chain types and $j$ is varied as $j=1, 2, 3......L$.  

\section{Results}\label{sec3}
\subsection*{(a) Correlation signal}
\begin{figure}[!thb]
\centerline{
\includegraphics[width = 3.3 in]{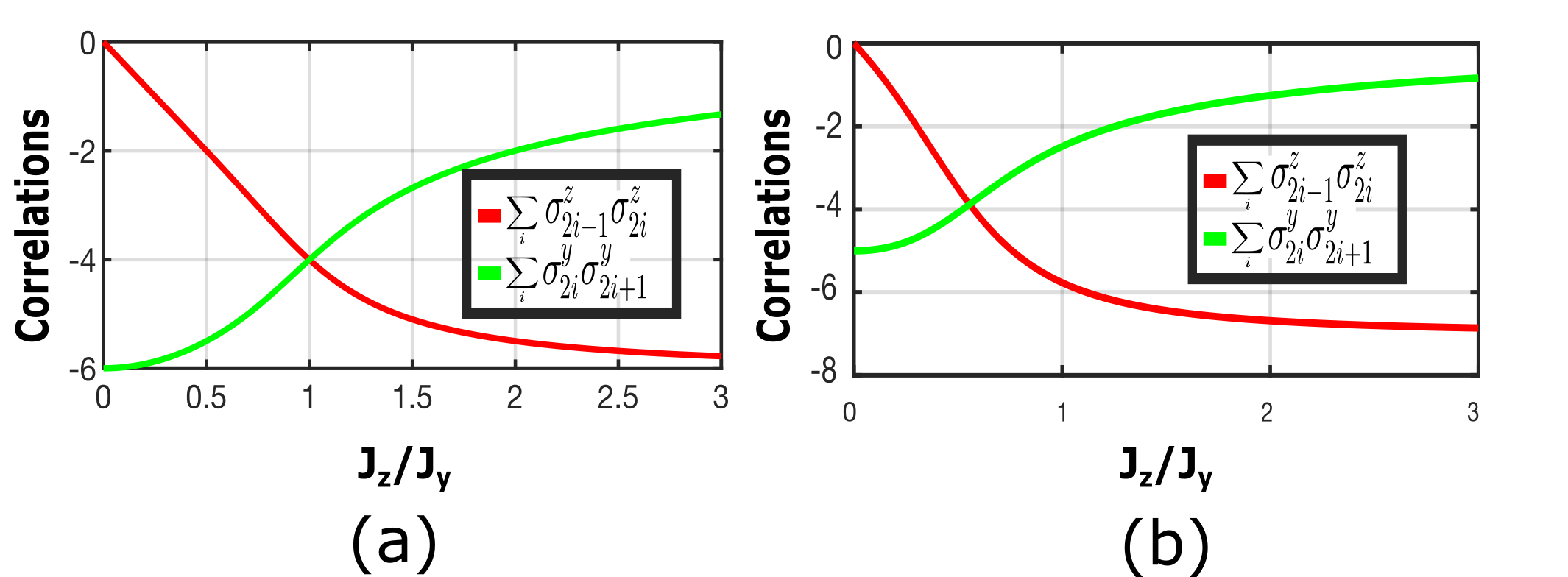}}
\caption{
Type 1:
(a) Correlation functions for defect at position 3 
(b) Correlation functions for defect at position 6 
}\label{fig:Figure_4}
\end{figure}

In case of type 1, we plot ZZ and YY correlation functions defined by Eqn. \ref{eq_corr} corresponding to two different qualitative situation. i.e. when the defect is at an odd site vs when the defect is at an even site. The ordering of ZZ versus YY is symmetric as is in the clean only when the number of zz bonds is equal to number of yy bonds. We find that QPT point at $J_z=J_y$ is shifted in the case of even defect site to $J_z=0.5\,J_y$  while it remains same for odd defect site. This is because for the former case, the total number of zz bonds and yy bonds are same and for the later case they are different. In general, we observe that the transition point is strongly susceptible to number of bonds and is proportional to the ratio of number of zz bonds over the number of yy bonds as shown in Fig. \ref{fig:Figure_4}. 
\par
\begin{figure}[!thb]
\centerline{
\includegraphics[width = 2 in]{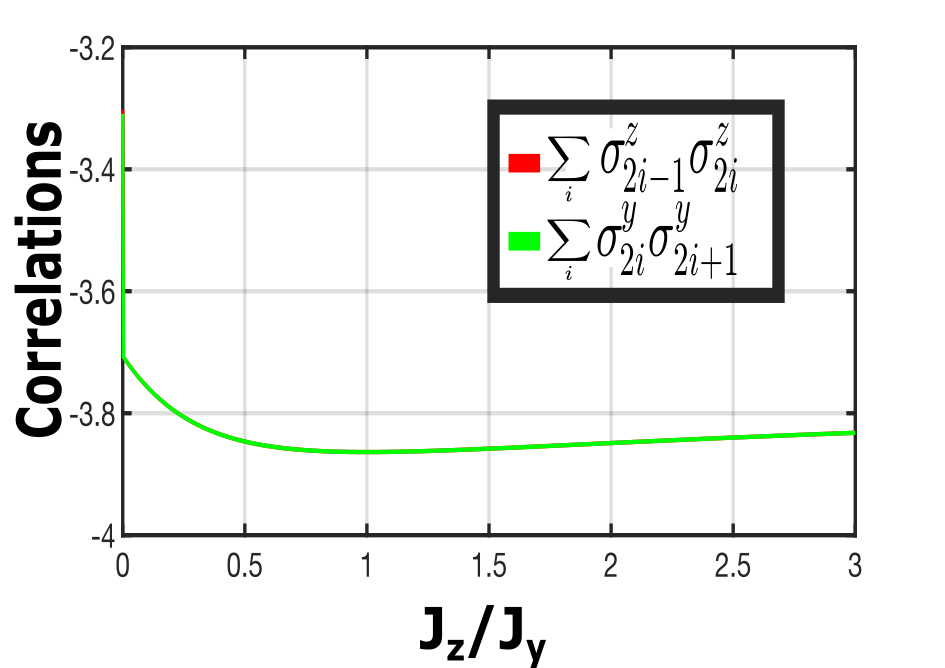}}
\caption{
Type 2:
Correlation functions for defect at position 4 with $J_z=J_y=1$. Both ZZ(red) and YY(green) correlations coincide with each other. 
}\label{fig:Figure_5}
\end{figure}

For type 2, we fix bond strengths of the bonds that are not defective to be $J_z=J_y=J$ and vary the bond strength of the defective bonds alone with $J_{\text{def}}$. The ZZ and YY correlations are plotted  as a function of defect strength fixing a defect site. We see that these correlations decay as we ramp the defect strength until $J_{\text{def}}=J$ post which it saturates as the entire contribution of these correlation quantities comes only from the defective bonds as shown in Fig. \ref{fig:Figure_5}.
\par
\begin{figure}[!thb]
\centerline{
\includegraphics[width = 3.3 in]{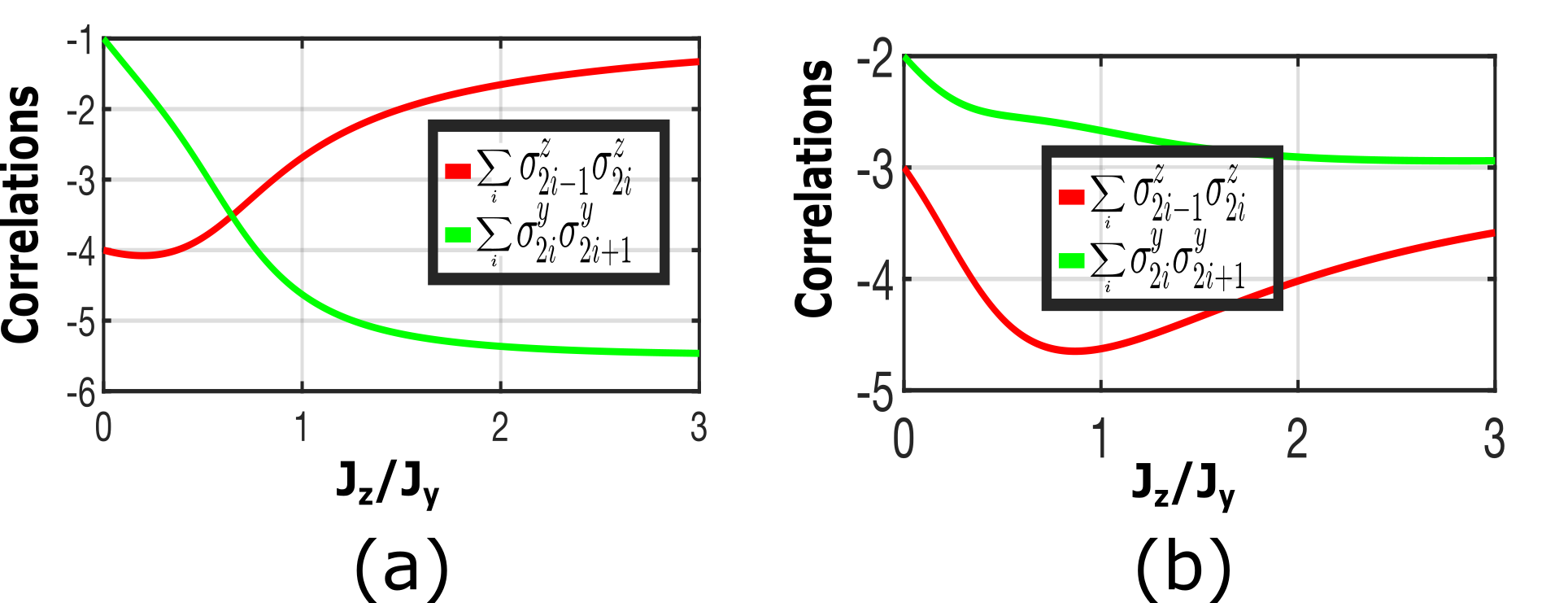}}
\caption{
Type 3: 
(a) Correlation functions for defect at position 3
(b) Correlation functions for defect at position 6 
}\label{fig:Figure_6}
\end{figure}

In case of type 3, we plot ZZ and YY correlation functions corresponding to two cases as taken in type 1. For the defect at site 3, The QPT at $J_z=J_y$ stays while interestingly for defect at site 6, although the ZZ and YY spin ordering follows a trend, the sharp QPT at $J_z=J_y$ is no more present. The behaviours of correlations is quite different from the case 1. For example, the ZZ correlations here are not zero when $J_z\rightarrow0$ as only the bond strengths are disordered while the alternating bond nature exists. This contributes to finite value of the correlations as long as $J_y\neq0$ as shown in Fig. \ref{fig:Figure_6}. 
\subsection*{(b) OTOC signal}
In this section, we plot the real and imaginary parts of the OTOC signal defined by Eq. \ref{eq_otoc} for 1D spin chain of length up to L=12 for all defect types.
\par
\begin{figure*}[!thb]
\centerline{
\includegraphics[width = 7 in]{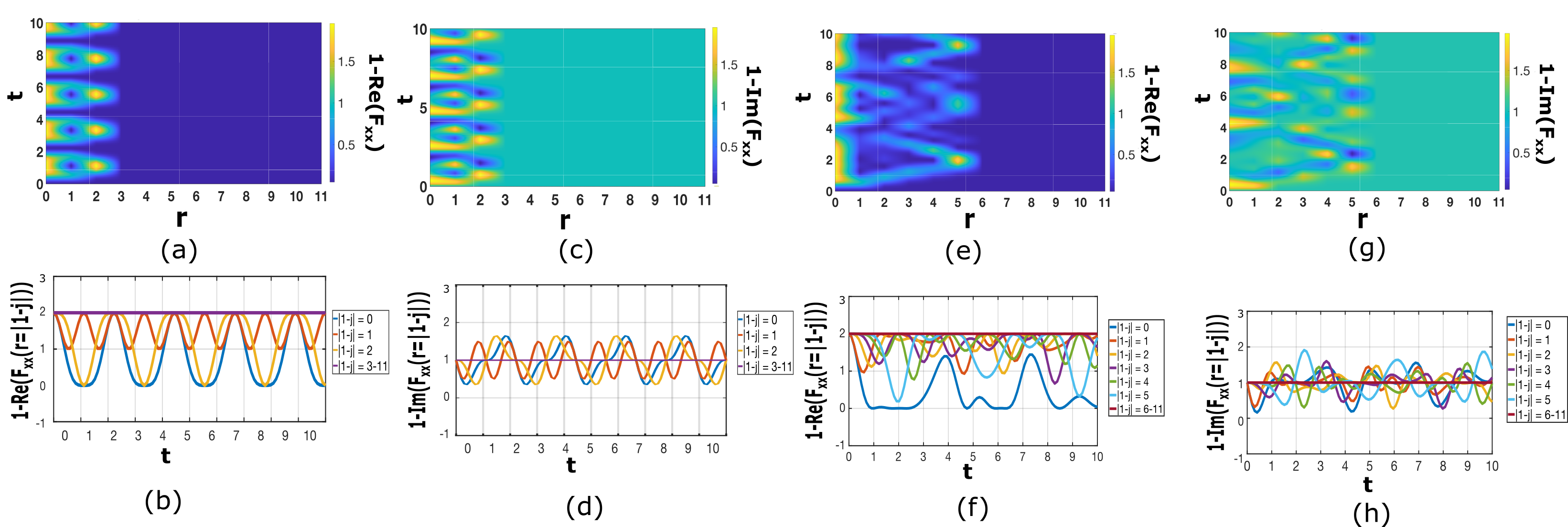}}
\caption{
Type 1:
(a) space-time propogation of real part of OTOC $F_{xx}(r,t)$ for defect at position 3
(b) Real part of individual OTOCs in time
(c) space-time propogation of imaginary part of OTOC $F_{xx}(r,t)$ for defect at position 3
(d) imaginary part of individual OTOCs in time
(e) space-time propogation of imaginary part of OTOC $F_{xx}(r,t)$ for defect at position 6
(f) Real part of individual OTOCs in time
(g) space-time propogation of imaginary part of OTOC $F_{xx}(r,t)$ for defect at position 6
(h) Imaginary part of individual OTOCs in time
}\label{fig:Figure_7}
\end{figure*}

 For type 1, the OTOC signal does not propagate beyond the defect site as evidentiated by numerical evidences in Figs. \ref{fig:Figure_7}. Both the real and imaginary parts of the considered XX-OTOC entirely capture the position of the defects. These observations can further be well understood and corroborated by BCH expansion of $\sigma_1^x(t)$. In the BCH expansion of $\sigma_1^x(t)$ (See Appendix \ref{app}), there are annihilator-like terms (Eqns. \ref{eq_annihilater_1a}-\ref{eq_annihilater_1b}) that prevent the growth of the operator's length beyond the defect site. Hence, the commutator which appears both in real and imaginary parts of the OTOC signal: $[\sigma_1^x(t),\sigma_{j}^x(0)]=0\,\text{for}\,j=d+1, d+2, ....n$ with $d=$defect position. This explains the observations in Figures \ref{fig:Figure_7} wherein the information propagation as well as scrambling is absent beyond the defect site thus giving rise to the phenomenon of Localization of information. 
\par
\begin{figure*}[!thb]
\centerline{
\includegraphics[width = 7 in]{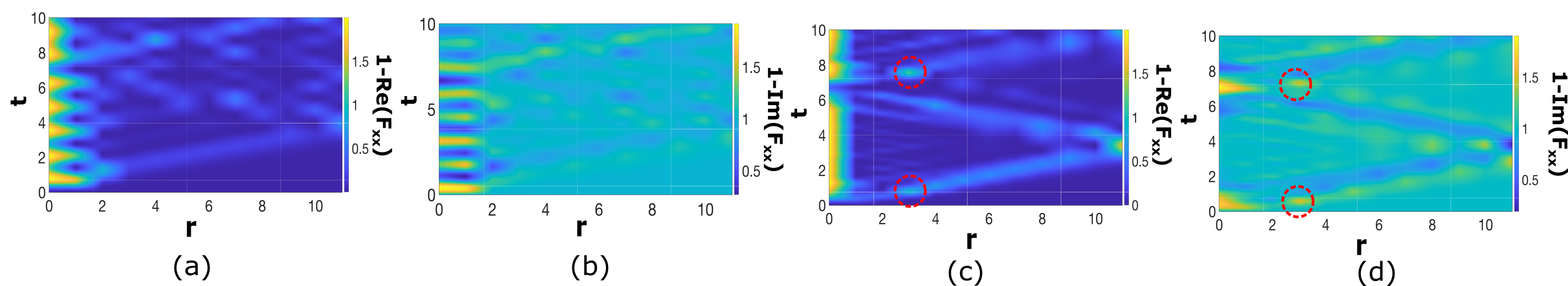}}
\caption{
Type 2, for $J_z=J_y=1$,
(a) space-time propogation of real part of OTOC $F_{xx}(r,t)$ for defect at position 3 for $J_\text{def}=0.5$ (defective bond strength)
(b) space-time propogation of imaginary part of OTOC $F_{xx}(r,t)$ for defect at position 3 for $J_\text{def}=0.5$ (defective bond strength)
(c) space-time propogation of real part of OTOC $F_{xx}(r,t)$ for defect at position 3 for $J_\text{def}=2$ (OTOC sign at the defect site marked in red)
(d) space-time propogation of imaginary part of OTOC $F_{xx}(r,t)$ for defect at position 3 for $J_\text{def}=2$ (OTOC sign at the defect site marked in red)
}\label{fig:Figure_8}
\end{figure*}

 For type 2, the OTOC signal propagates giving signs of the defect as the bond strength of the defective bonds $J_{\text{def}}$ is increased as in Fig. \ref{fig:Figure_8}. Both the real and imaginary parts of the considered XX-OTOC captures the defects(marked in red). The reason for these observations is understood by the BCH expansion of $\sigma_1^x(t)$. In the BCH expansion of $\sigma_1^x(t)$, the annihilator-like terms as in \ref{eq_annihilater_2} for type 2 defects, has an amplitude or weight that depends on the defective bond strength $J_{\text{def}}$. This explains the observations in Figures \ref{fig:Figure_8} wherein the information propagation as well as scrambling is localized in the illustrated fashion giving necessary signs of the defects. 

\par
\begin{figure*}[!thb]
\centerline{
\includegraphics[width = 7 in]{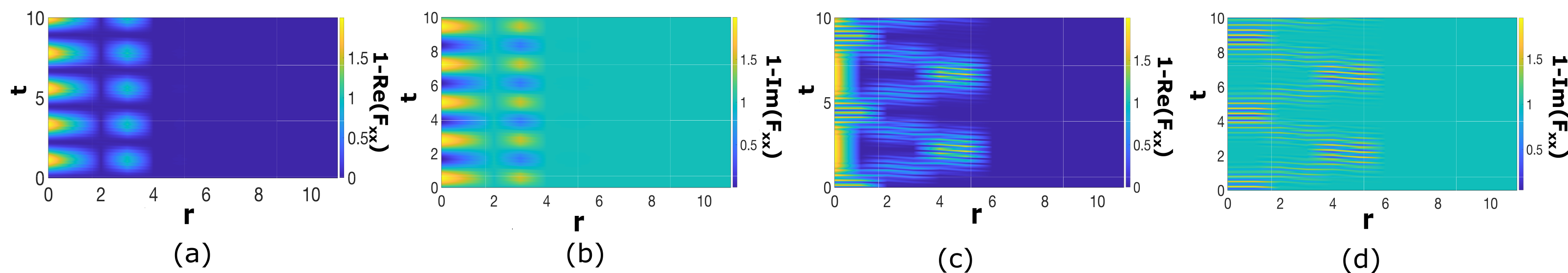}}
\caption{
Type 3, 
(a) space-time propogation of real part of OTOC $F_{xx}(r,t)$ for defect at position 3 with $J_y=10,\, Jz=.1$
(b) space-time propogation of imaginary part of OTOC $F_{xx}(r,t)$ for defect at position 3 with $J_y=10,\, Jz=.1$
(c) space-time propogation of real part of OTOC $F_{xx}(r,t)$ for defect at position 6 with $J_y=.1,\,Jz=1$
(d) space-time propogation of imaginary part of OTOC $F_{xx}(r,t)$ for defect at position 6 with $J_y=.1,\,Jz=1$
}\label{fig:Figure_9}
\end{figure*}

 For type 3, the OTOC signal does not propagate beyond the defect site as shown in Fig. \ref{fig:Figure_9}. Similar to the above cases, both the real and imaginary parts of the considered XX-OTOC capture defect signatures. The annihilator-like terms (Eqns. \ref{eq_annihilater_3a}-\ref{eq_annihilater_3b}) in the BCH expansion of $\sigma_1^x(t)$ has an amplitude that depends on the ratio $\frac{1}{J_{y}}$ for an odd defect site and ratio $\frac{1}{J_{z}}$ for even defect site. By tuning this ratio, OTOC signal can be made localized thus giving suitable signs of the defects.

\section{Conclusion}\label{sec4}
We have studied defects on pristine 1D Kitaev model using structural and dynamical quantities. We have presented ways to realize three types of disorders that are prevalent in Kitaev materials.  In particular, the effects of disorder on the spin-spin correlations and OTOCs within the ground-state manifold of our defective models have been investigated. We have considered 3 types of disorder: bond nature-flip, bond-strength, bond strength-flip disorders. Though these disorders are quite different, they behave similar in terms of their responses to the OTOC signal probe i.e. they cause localization of information in the OTOC space thus illustrating prohibited information scrambling across the length of the spin-chain. Regardless of this localization phenomena, the disorders show quite unique signatures of themselves in the OTOC space. This makes OTOCs as suitable detection tools susceptible to different defects in the model. In terms of physical realization, not only can OTOCs be measured in an experimental setup\cite{qc5}, circuit-based measurements of OTOCs on state-of the art quantum computers have also been acheived\cite{qc1,qc2,qc3,qc4}. Moreover, in superconducting Circuit quantum electrodynamics(CQED) setting\cite{cqed}, the above-discussed defective models can be realized using superconducting circuits\cite{qc2,chavez2023spectral} wherein the defective qubit along with its local bonds can be used as a control qubit that controls the entire scrambling dynamics of the circuit. 

\par Though we have considered only time-independent and static disorders, the OTOC signal proposed in this article, given its spatio-temporal dependence can be used to detect time-dependent perturbations too upon the clean model. A special consideration is floquet or periodic time-dependent perturbations wherein the relevant quantity to consider is the Floquet OTOC\cite{floquet_dpt,floq_2,floq_3}. Such systems have shown to exhibit interesting phases such as time crystals and MBL \cite{tc_1,tc_2,tc_3,tc_4}. These reasons serve as motivating factors for a future study of these phases in 1D kitaev model. Additionally, as a natural extension, study of OTOC propogation in 2D Kitaev model can give insights into defects and fundamental excitations.
\acknowledgments
The authors would like to acknowledge the financial support from the Quantum Science Center, a National Quantum Information Science Research Center of the U.S. Department of Energy (DOE).
\newpage
\input{manuscript.bbl}
\appendix
\begin{widetext}
\section{Time Evolution of $\sigma_1^x(t)$}\label{app}

The time evolution of $\sigma_1^x(t)$ is given by Heisenberg time evolution which can be further expanded with the help of Baker-Campbell-Haunsdroff(BCH)\cite{hall2013lie} formula as,

\begin{align}
\sigma_1^x(t)=&e^{iHt}\sigma_1^x(0)e^{-iHt} \nonumber \\
&=\sigma_1^x(0)+it[H,\sigma_1^x]+\frac{(it)^2}{2!}[H,[H,\sigma_1^x]]+.......\frac{(it)^n}{n!}[H,\sigma_1^x]_{n}
\end{align}
where $[H,\sigma_1^x(t)]_{n}$ is the nested commutator obtained after computing $[H,[H,....\text{n times....}[H,\sigma_1^x(t)]$.
\subsection*{Type 1}
\begin{itemize}
    \item \textbf{\underline{\textit{case 1}}}: When defect occurs at an odd site, the defective bond is a y-y bond, thus resulting in the hamiltonian,
    \begin{align}
     H=&J(\sigma_{1}^z \sigma_{2}^z+\sigma_{2}^y \sigma_{3}^y+\sigma_{3}^z \sigma_{4}^z+.............+\sigma_{d-2}^z \sigma_{d-1}^z+\underbrace{\sigma_{d-1}^y \sigma_{d}^y+\sigma_{d}^y \sigma_{d+1}^y}_\text{defective bond}+\sigma_{d+1}^z \sigma_{d+2}^z+\sigma_{d+2}^y \sigma_{d+3}^y+...........\nonumber 
    \end{align}
where d=defect position. 
$\sigma_1^x(t)$ under the evolution of the above Hamiltonian has the form,
\begin{align}
\sigma_1^x(t)=\sigma_1^x(0)+it(J2i\sigma_1^y\sigma_2^z)........\underbrace{+\frac{(it)^{d-1}}{(d-1)!}2^{d-1}i^{d-1}J^{d-1}\sigma_1^y\sigma_1^x.....\sigma_{d-1}^x\sigma_{d}^y}_\text{Annihilator} \label{eq_annihilater_1a}
\end{align}
  \item \textbf{\underline{\textit{case 2}}}: When defect occurs at an even site, the defective bond is a z-z bond, thus resulting in the hamiltonian,
    \begin{align}
     H=&J(\sigma_{1}^z \sigma_{2}^z+\sigma_{2}^y \sigma_{3}^y+\sigma_{3}^z \sigma_{4}^z+.............+\sigma_{d-2}^y \sigma_{d-1}^y+\underbrace{\sigma_{d-1}^z \sigma_{d}^z+\sigma_{d}^z \sigma_{d+1}^z}_\text{defective bond}+\sigma_{d+1}^y \sigma_{d+2}^y+\sigma_{d+2}^z \sigma_{d+3}^z+...........\nonumber 
    \end{align}
where d=defect position. 
$\sigma_1^x(t)$ under the evolution of the above Hamiltonian has the form,
\begin{align}
\sigma_1^x(t)=\sigma_1^x(0)+it(J2i\sigma_1^y\sigma_2^z)........\underbrace{+\frac{(it)^{d-1}}{(d-1)!}2^{d-1}i^{d-1}J^{d-1}\sigma_1^y\sigma_1^x.....\sigma_{d-1}^x\sigma_{d}^z}_\text{Annihilator} \label{eq_annihilater_1b}
\end{align}
\end{itemize}
The operator of the annihilator terms in the BCH expansion gets terminated with $\sigma_{d}^z(\sigma_{d}^y)$ for odd d(even d) correspondingly that prohibits their growth. This in turn, results in the commutator: $[\sigma_1^x(t),\sigma_{j}^x(0)]=0\,\text{for}\,j\geq d+1$ with $d=$defect position.  
\subsection*{Type 2}
When defect occurs at a site d, it results in the hamiltonian,
    \begin{align}
     H=&J(\sigma_{1}^z \sigma_{2}^z+\sigma_{2}^y \sigma_{3}^y+\sigma_{3}^z \sigma_{4}^z+.............+\sigma_{d-2}^z \sigma_{d-1}^z)+J_{\text{def}}(\underbrace{\sigma_{d-1}^y \sigma_{d}^y+\sigma_{d}^z \sigma_{d+1}^z}_\text{defect})+J(\sigma_{d+1}^z \sigma_{d+2}^z+\sigma_{d+2}^y \sigma_{d+3}^y+...........\nonumber 
    \end{align}
where d=defect position. 
$\sigma_1^x(t)$ under the evolution of the above Hamiltonian has the form,
\begin{align}
\sigma_1^x(t)=\sigma_1^x(0)+it(J2i\sigma_1^y\sigma_2^z)........\underbrace{+\frac{(it)^{d-1}}{(d-1)!}2^{d-1}i^{d-1}\frac{J^{d}}{J_{\text{def}}}\sigma_1^y\sigma_1^x.....\sigma_{d-1}^x\sigma_{d}^x}_\text{Annihilator}+........\label{eq_annihilater_2}
\end{align}
As the bond strength $J_{\text{def}}$ is gradually ramped up, the weight of the operator $\sigma_1^y\sigma_1^x.....\sigma_{d-1}^x\sigma_{d}^x$ dies down due to which the subsequent nested commutators in the BCH expansion is heavily attenuated giving rise to OTOC localization as illustrated in the main text of this article.
\subsection*{Type 3}
\begin{itemize}
    \item \textbf{\underline{\textit{case 1}}}: When defect occurs at an odd site, it results in the hamiltonian,
    \begin{align}
     H=&J_z\sigma_{1}^z \sigma_{2}^z+J_y\sigma_{2}^y \sigma_{3}^y+\sigma_{3}^z \sigma_{4}^z+.............+\sigma_{d-2}^z \sigma_{d-1}^z+J_y\underbrace{\sigma_{d-1}^y \sigma_{d}^y+J_y\sigma_{d}^z \sigma_{d+1}^z}_\text{defect}+J_z\sigma_{d+1}^y \sigma_{d+2}^y+J_y\sigma_{d+2}^z \sigma_{d+3}^z+...........\nonumber 
    \end{align}
where d=defect position. 
$\sigma_1^x(t)$ under the evolution of the above Hamiltonian has the form,
\begin{align}
\sigma_1^x(t)=\sigma_1^x(0)+it(J_z2i\sigma_1^y\sigma_2^z)........\underbrace{+\frac{(it)^{d-1}}{(d-1)!}2^{d-1}i^{d-1}\frac{(J_zJ_y)^d}{J_y}\sigma_1^y\sigma_1^x.....\sigma_{d-1}^x\sigma_{d}^y}_\text{Annihilator}  \label{eq_annihilater_3a}
\end{align}
  \item \textbf{\underline{\textit{case 2}}}: When defect occurs at an even site, it results in the hamiltonian,
    \begin{align}
     H=&J_z\sigma_{1}^z \sigma_{2}^z+J_y\sigma_{2}^y \sigma_{3}^y+\sigma_{3}^z \sigma_{4}^z+.............+\sigma_{d-2}^y \sigma_{d-1}^y+J_z\underbrace{\sigma_{d-1}^z \sigma_{d}^z+J_z\sigma_{d}^y \sigma_{d+1}^y}_\text{defect}+J_y\sigma_{d+1}^z \sigma_{d+2}^z+J_z\sigma_{d+2}^y \sigma_{d+3}^y+...........\nonumber 
    \end{align}
where d=defect position. 
$\sigma_1^x(t)$ under the evolution of the above Hamiltonian has the form,
\begin{align}
\sigma_1^x(t)=\sigma_1^x(0)+it(J_z2i\sigma_1^y\sigma_2^z)........\underbrace{+\frac{(it)^{d-1}}{(d-1)!}2^{d-1}i^{d-1}\frac{(J_zJ_y)^d}{J_z}\sigma_1^y\sigma_1^x.....\sigma_{d-1}^x\sigma_{d}^z}_\text{Annihilator} \label{eq_annihilater_3b}
\end{align}
\end{itemize}

\end{widetext}
\end{document}

%% file: manuscript.bbl
%